\documentclass[preprintnumbers,amsmath,amssymb,letterpaper,nofootinbib,notitlepage]{revtex4}
\usepackage{graphicx}
\usepackage{enumerate}
\usepackage{bm}
\usepackage{slashed}

\newcommand\Eq[1]{Eq.~\ref{eq:#1}}
\newcommand\Fig[1]{Fig.~\ref{fig:#1}}

\usepackage{amssymb}
\usepackage{amsmath}
\usepackage{epsfig}
\newcommand{\be}{\begin{equation}}
\newcommand{\ee}{\end{equation}}
\newcommand\beq{\begin{eqnarray}}
\newcommand\eeq{\end{eqnarray}}

\newcommand{\Tr}{{\rm Tr\,}}
\newcommand{\tr}{{\rm tr\,}}

\newcommand{\threeptwo}{${}^3P_2$ }
\newcommand{\lrnabla}{\overleftrightarrow\nabla}
\newcommand{\nn}{\nonumber}

\newcommand{\mybar}[1]%
        {\kern 0.6pt\overline{\kern -0.6pt#1\kern -0.6pt}\kern 0.6pt}
\begin{document}

\preprint{UM-DOE/ER/40762-527 }

\title{Massive and massless modes of the triplet phase of neutron matter}

\author{Paulo F. Bedaque}
\email{bedaque@umd.edu}

\author{Amy N. Nicholson}
 \email{amynn@umd.edu}
 
 \author{Srimoyee Sen}
 \email{srimoyee@umd.edu}
  
\affiliation{Maryland Center for Fundamental Physics, Department of Physics,
University of Maryland, College Park MD 20742-4111, USA}


 
 \begin{abstract}
Neutron matter at densities of the order of the nuclear saturation density is believed to have neutrons paired in the \threeptwo channel. We study the low lying modes of this phase and find two massless modes (angulons), resulting from the spontaneous breaking of rotational symmetry as well as three other, gapped modes. We compute their masses at arbitrary temperatures.
\end{abstract}

\maketitle

\section{Introduction}
Understanding the fate of matter at densities comparable or even larger than nuclear densities, as found in neutron stars, is one of the outstanding open problems in Physics and Astrophysics. Due to our current lack of understanding of non-perturbative QCD the problem is generally approached by a combination of model/effective theory calculations and neutron star observations. The equation of state of matter at these high densities has a one-to-one correspondence with the relation between the mass and the radius of a neutron star. Consequently, substantial effort has been invested in answering the difficult question of how to  determine masses and radii of neutron stars. Even if this program is successful, it can only provide information about the (essentially zero temperature) equation of state. As interesting as this is, qualitatively different phases of matter can have very similar equations of state. In order to distinguish different phases more subtle observables are required. A whole class of observables related to transport properties and finite temperature effects (heat and charge conductivities, neutrino opacities, specific heat) are related to measurable properties of the star (cooling curves, R-mode stability). Low temperature transport properties are determined by the low lying excitations of the ground state, so the theoretical understanding of their properties is an essential step towards connecting a particular putative phase of dense matter with neutron star observations.

Since the early days of pulsar physics it was recognized that at densities equal to or higher than nuclear saturation densities, neutrons near the Fermi surface should form Cooper pairs in the \threeptwo channel, and since then, model calculations have strengthened  that belief. An intuitive way to understand this is to observe that the neutron-neutron phase shifts become repulsive at momenta comparable to the Fermi momentum in dense matter in all but the \threeptwo channel. Attraction at the Fermi surface, no matter how small, generically leads to Cooper pair formation and superfluidity. In fact, the process of formation of \threeptwo pairs and their breaking (with the concomitant emission of a neutrino pair) was invoked recently to explain the rapid cooling observed in Cassiopeia A. 

It is thus important to understand the low lying excitations of the \threeptwo superfuild phase of neutron matter.
It was pointed out in \cite{Bedaque:2003wj} that since the formation of triplet Cooper pairs breaks rotational symmetry,  one would expect that gapless Goldstone modes, named ``angulons", should exist. An effective theory for the angulons was obtained on very general grounds in \cite{Bedaque:2012bs} and their consequences to transport was discussed in \cite{Bedaque:2013fja,Bedaque:2013rya}. All of these discussions hinge on the validity of the Goldstone theorem, however, there are reasons to proceed more carefully in this case because the Goldstone theorem is more subtle when {\it  spacetime } symmetries are broken.  In addition, a model calculation (\cite{Leinson:2012pn}) has found that the excitations  that are gapless at zero temperature develop a mass at finite temperatures. 
There is also the possibility that some of the gapped modes, ignored in previous analyses, are so light as to contribute to transport properties at the low temperatures of a neutron star.
The main goal of this paper is to discuss these questions through the analysis of a simple model containing the essential symmetry properties of a realistic calculation.

\section{Action for the fluctuations around the ground state}

We begin with a microscopic model describing non-relativistic fermions with an attractive contact interaction of the form,
\beq\label{eq:L}
\mathcal{L} = \psi^{\dagger}\left(i\partial_0-\epsilon_{(-i\nabla)}\right)\psi -\frac{g^2}{4}\left(\psi^{\dagger}\sigma_i\sigma_2\lrnabla_j\psi^{*}\right)
\chi^{kl}_{ij}
\left(\psi^T\sigma_2\sigma_k\lrnabla_l\psi\right) \ ,
\eeq 
where $\chi^{kl}_{ij}= \frac{1}{2}(\delta_{ik}\delta_{jl}+\delta_{il}\delta_{jk}-\frac{2}{3}\delta_{ij}\delta_{kl})$ is the projector onto the ${}^3P_2$ state,  $\epsilon_p = \sqrt{\frac{p^2}{2M} - \mu}$ is the kinetic energy of fermonic quasiparticles near the Fermi surface having neutron quantum numbers, and $M$ and $\mu$ are the fermion mass and chemical potential, respectively (see \cite{Bedaque:2012bs} for further discussion of the interpretation of the model in terms of a Fermi liquid theory). In order to calculate finite temperature results, the calculation will proceed in imaginary time, however, to find the normal modes we must later analytically continue back to Minkowski space.

Below the threshold for the breakup and formation of Cooper pairs, we expect the low-lying modes to consist of bosonic fluctuations of the order parameter. Thus, we introduce an auxiliary field, $\Delta_{ij} = \Delta_0 +\delta \Delta(x)$, where $\Delta_0 = \langle n^T\sigma_2 \sigma_i \lrnabla_j n\rangle$ is the value of the order parameter and $\delta \Delta(x)$ encodes fluctuations about the ground state, resulting in,
\beq
S =\int d^4x\left[\psi^{\dagger}(i\partial_0-\epsilon(-i\nabla))\psi+\frac{1}{4g^2}\Delta^{\dagger}_{ij}\Delta_{ji}+\frac{\Delta^{\dagger}_{ij}}{4}(\psi^T\sigma_2\sigma_i\lrnabla_j\psi)-\frac{\Delta_{ji}}{4}(\psi^\dagger\sigma_i\sigma_2\lrnabla_j\psi^{*}) \right]\ .
\eeq The presence of the projector $\chi^{kl}_{ij}$ is unnecessary in the expression above as we will consider only traceless, symmetric $\Delta_{ij}$.
As discussed in \cite{Bedaque:2012bs}, there are several possible $^3P_2$ phases governed by the symmetry breaking pattern of the order parameter, a symmetric traceless tensor. The phase corresponding to the ground state of relevance for neutron stars is unclear, but there is some evidence that the condensate should be a real matrix \cite{Richardson:1972xn}, and that a condensate of the form,
\beq
\label{eq:groundState}
\Delta_0&=\bar{\Delta}\left(\begin{array}{ccc}
-1/2&0&0 \\
0&-1/2&0 \\
0&0&1 \\
\end{array}\right) \ ,
\eeq
where $\bar{\Delta}$ is the magnitude of the gap, may be favored \cite{Vulovic:1984kc}. In this phase, rotational invariance in one plane is maintained, and two massless Goldstone modes, corresponding to the breaking of rotational invariance in the remaining two planes, are expected. Throughout the calculation we will remain as general as possible regarding which phase the system is in, however, we do assume that $\Delta$ is a real symmetric matrix\footnote{Note that if $\Delta$ is assumed to be a real matrix it will not describe the usual $U(1)$ superfluid phonon associated with broken baryon number.} and will specialize to the above phase before presenting results. 

We now integrate out the fermion fields and convert to momentum space to obtain,
\beq
S[\Delta]=-T\int \frac{d^3p}{(2\pi)^3}\sum_{p_0}\left(\frac{1}{4g^2}\Delta^{\dagger}_{ij}(p)\Delta_{ji}(p)\right)+\text{Tr}\log(D^{-1})\ ,
\label{eqn4}
\eeq
where the kernel of $D^{-1}$ is
\beq
D^{-1}(p,k)=\begin{pmatrix}
(2\pi)^3\delta^3(\mathbf{p}-\mathbf{k})\delta_{p_0,k_0}(ip_0+\epsilon_p) && i\Delta_{ji}(\mathbf{k}-\mathbf{p})\sigma_i\sigma_2p_j \cr
-i\Delta_{ij}(-\mathbf{k}+\mathbf{p})\sigma_2\sigma_ip_j && (2\pi)^3\delta^3(\mathbf{p}-\mathbf{k})\delta_{p_0,k_0}(ip_0-\epsilon_p)
\end{pmatrix}\ .
\eeq
To study small fluctuations of the field we expand the action, $S[\Delta]=S[\Delta_0]+\frac{\partial S}{\partial \Delta}\bigr\rvert_{\Delta=\Delta_0}\delta\Delta+\frac{1}{2}\frac{\partial^2S}{\partial\Delta^2}\bigr\rvert_{\Delta=\Delta_0}(\delta\Delta)^2+...$. $\Delta_0$ satisfies the gap equation, $\frac{\partial S}{\partial \Delta}\bigr\rvert_{\Delta=\Delta_0}=0$, so that the first term of interest is quadratic in $\delta\Delta$.

Using 
\beq
\left.\frac{\delta^2\Tr\log(D^{-1})}{\delta\Delta_{ij}(s)\delta\Delta_{kl}(r)}\right|_{\Delta=\Delta_0}=\Tr\left(D(p,q')\frac{\delta D^{-1}(q',q)}{\delta\Delta_{ij}(s)}D(q,p')\frac{\delta D^{-1}(p',k)}{\delta\Delta_{kl}(r)}\right)
\eeq 
where `Tr' denotes the trace in Gorkov, spin, and momentum space, we find,
\beq
\left.\frac{\delta^2\Tr\log(D^{-1})}{\delta\Delta_{ij}(s)\delta\Delta_{kl}(r)}\right.|_{\Delta=\Delta_0}&=& T \sum_{p_0}\int \frac{d^3p}{(2\pi)^3}\tr\left[\frac{\sigma^m\sigma^i\sigma^{m'}\sigma^k(\mathbf{p}-\mathbf{r})^j(\mathbf{p}-\mathbf{r})^{n'}p^lp^n\left(\Delta_0\right)_{mn}\left(\Delta_0\right)_{mn'}}{(p_0^2+E_{\mathbf{p}}^2)((p_0-\omega)^2+E_{\mathbf{p-r}}^2)}\right.\cr
&+& \left.  \frac{\sigma^i\sigma^l(\mathbf{p}-\mathbf{r})^j\mathbf{p}^k(ip_0+\epsilon_{\mathbf{p}})(i(p_0-\omega)-\epsilon_{\mathbf{p}-\mathbf{r}})}{(p_0^2+E_{\mathbf{p}}^2)((p_0-\omega)^2+E_{\mathbf{p-r}}^2)}+ \frac{\sigma^j\sigma^k(\mathbf{p}+\mathbf{r})^i\mathbf{p}^l(ip_0-\epsilon_{\mathbf{p}})(i(p_0+\omega)+\epsilon_{\mathbf{p}+\mathbf{r}})}{(p_0^2+E_{\mathbf{p}}^2)((p_0+\omega)^2+E_{\mathbf{p+r}}^2)} \right.\cr
&+&\left. \frac{\sigma^m\sigma^j\sigma^{m'}\sigma^l(\mathbf{p}+\mathbf{r})^i(\mathbf{p}+\mathbf{r})^{n'}p^kp^n\left(\Delta_0\right)_{mn}\left(\Delta_0\right)_{m'n'}}{(p_0^2+E_{\mathbf{p}}^2)((p_0+\omega)^2+E_{\mathbf{p+r}}^2)}\right] \delta^3(\mathbf{s}+\mathbf{r})\frac{\delta_{r_0+s_0,0}}{T}
\eeq
where $\omega \equiv r_0$, $\sum_{p_0} f(p_0)\equiv \sum_{n=-\infty}^{\infty} f((2n+1)\pi T)$, $E_{\mathbf{p}}^2 \equiv \epsilon_{\mathbf{p}}^2+\mathbf{p}\cdot\Delta_0\cdot\Delta_0\cdot\mathbf{p}$, and the remaining trace is over spin indices only. In this work we will not compute dispersion relations and will therefore restrict our calculation to zero spatial momentum. After completing the traces we find the simplified form,

\beq
\frac{\delta^2\Tr\log(D^{-1})}{\delta\Delta_{ij}(s)\delta\Delta_{kl}(r)}=T\sum_{p_0}\int \frac{d^3p}{(2\pi)^3}\left[\frac{8\mathbf{p}_i[\Delta_0\cdot \mathbf{p}]_j\mathbf{p}_k[\Delta_0\cdot \mathbf{p}]_l-4\mathbf{p}_i \mathbf{p}_l\left(p_0(p_0+\omega)+E_{\mathbf{p}}^2\right)}{(p_0^2+E_{\mathbf{p}}^2)\left((p_0+\omega)^2+E_{\mathbf{p}^2}\right)}\right] \delta^3(\mathbf{s}+\mathbf{r})\frac{\delta_{r_0+s_0,0}}{T}
\eeq

The second order action in the $\delta \Delta$ expansion is,
\beq
\label{eq:S2presum}
S_2 &=& T\sum_{p_0}\int \frac{d^3p}{(2\pi)^3}  \left[\frac{4(\mathbf{p}\cdot \delta\Delta\cdot\Delta_0\cdot \mathbf{p})^2+\left(\omega^2-2(p_0^2+E_{\mathbf{p}}^2)\right)(\mathbf{p}\cdot \delta\Delta\cdot\delta\Delta\cdot \mathbf{p})}{(p_0^2+E_{\mathbf{p}}^2)\left((p_0+\omega)^2+E_{\mathbf{p}}^2\right)}+ \frac{\frac{4}{3}\Tr [\delta \Delta \cdot \delta \Delta] (\mathbf{p}\cdot \hat{\Delta}_0\cdot\hat{\Delta}_0\cdot \mathbf{p})}{p_0^2+E_{\mathbf{p}}^2}\right] \ ,
\eeq
where $\hat{\Delta}_0 \equiv \Delta_0/\bar{\Delta}$ and we have used the gap equation (see Appendix~\ref{sec:gap}) to substitute the coupling for the magnitude of the gap. We may now perform the sum over $p_0$. We simplify our expression by setting $\omega = 2\pi m T  \ (m \in\mathbb{Z}$, corresponding to bosonic modes) within all trigonometric functions. This is also necessary to give us the correct analytic continuation to Minkowski space (see Appendix~\ref{sec:realTime}). The result for $\omega\neq 0$ is given by
\beq
\label{eq:S2postSum}
S_2 =\int \frac{d^3p}{(2\pi)^3}\tanh\left(\frac{E_{\mathbf{p}}}{2T}\right)\left[\frac{4(\mathbf{p}\cdot \delta\Delta\cdot\Delta_0\cdot \mathbf{p})^2-4E_{\mathbf{p}}^2(\mathbf{p}\cdot \delta\Delta\cdot\delta\Delta\cdot \mathbf{p})}{E_{\mathbf{p}}\left(\omega^2+4E_{\mathbf{p}}^2\right)} + \frac{\frac{2}{3}\Tr [\delta \Delta \cdot \delta \Delta]  (\mathbf{p}\cdot \hat{\Delta}_0\cdot\hat{\Delta}_0\cdot \mathbf{p})}{E_{\mathbf{p}}}\right]\ .
\eeq

\section{Normal modes}

An arbitrary 3-dimensional symmetric, traceless matrix may be expanded in an orthonormal basis of the following matrices:
\beq
\label{eq:orthBasis}
\mathcal{M}^{(1)} &\equiv & \left( \begin{array}{ccc} 
0 & 0 & 0 \\
0 & 0 & 1 \\
0 & 1 & 0 \\
\end{array} \right)  \ , \qquad \mathcal{M}^{(2)} \equiv \left( \begin{array}{ccc} 
0 & 0 & 1 \\
0 & 0 & 0 \\
1 & 0 & 0 \\
\end{array} \right) \ , \qquad \mathcal{M}^{(3)} \equiv \left( \begin{array}{ccc} 
0 & 1 & 0 \\
1 & 0 & 0 \\
0 & 0 & 0 \\
\end{array} \right) \cr
&&\mathcal{M}^{(4)} \equiv  \left( \begin{array}{ccc} 
1 & 0 & 0 \\
0 & -1 & 0 \\
0 & 0 & 0 \\
\end{array} \right)  \ , \qquad \mathcal{M}^{(5)} \equiv \left( \begin{array}{ccc} 
-1/2 & 0 & 0 \\
0 & -1/2 & 0 \\
0 & 0 & 1 \\
\end{array} \right) \ . 
\eeq
It is simple to show that, given our chosen ground state (\Eq{groundState}), the kinetic and potential terms induce no mixing between states corresponding to these matrices, thus, they correspond to eigenstates of the Hamiltonian operator. To do so, we define the operator, 
\beq
\mathcal{H}_{ij}[\delta \Delta] &\equiv& \frac{\delta S_2[\delta \Delta]}{\delta \delta \Delta_{ij}} = \int \frac{d^3p}{(2\pi)^3}\left[ \frac{\tanh\left(\frac{E_{\mathbf{p}}}{2T}\right) }{E_{\mathbf{p}}\left(\omega^2+4E_{\mathbf{p}}^2\right)}\left( 8 \mathbf{p}\cdot \delta\Delta\cdot\Delta_0\cdot \mathbf{p} \left(\mathbf{p}_i\left[\Delta_0\right]_{jk} \mathbf{p}_k +\mathbf{p}_j\left[\Delta_0\right]_{ik} \mathbf{p}_k\right) \right. \right. \cr
&-& \left.\left. 8E_{\mathbf{p}}^2\left(\mathbf{p}_i\delta\Delta_{jk} \mathbf{p}_k+\mathbf{p}_j\delta\Delta_{ik} \mathbf{p}_k\right)+ \frac{4}{3}\left(\omega^2+4E_{\mathbf{p}}^2\right) \left(\delta\Delta_{ij}+\delta\Delta_{ji}\right) (\mathbf{p}\cdot \hat{\Delta}_0\cdot\hat{\Delta}_0\cdot \mathbf{p})\right) \right] \ ,
\eeq
where we have used the fact that $\delta \Delta$ is symmetric in taking the derivative. Then, by orthonormality of the set of matrices, if $\mathcal{H}[\delta \Delta] \propto \delta \Delta$ for all matrices in \Eq{orthBasis}, there can be no mixing between states. As an example, if we choose $\delta \Delta = \bar{\Delta} \mathcal{M}^{(1)}$, then we have
\beq
&&\mathcal{H}[\mathcal{M}^{(1)}] = \int \frac{d^3p}{(2\pi)^3}\frac{2\bar{\Delta}\tanh\left(\frac{E_{\mathbf{p}}}{2T}\right)}{E_{\mathbf{p}}\left(\omega^2+4E_{\mathbf{p}}^2\right)} \left( \begin{array}{ccc}
-2p_x^2 p_y p_z \bar{\Delta}^2 & -2 \bar{\Delta}^2 p_x p_y^2 p_z-4E_{\mathbf{p}}^2p_x p_z & \bar{\Delta}^2 p_x p_y p_z^2-4E_{\mathbf{p}}^2p_x p_y \cr
-2\bar{\Delta}^2p_x p_y^2 p_z-4E_{\mathbf{p}}^2p_x p_z & -2 \bar{\Delta}^2 p_y^3 p_z-8E_{\mathbf{p}}^2p_y p_z & \begin{array}{c}
\scriptstyle \bar{\Delta}^2 p_y^2 p_z^2+ \frac{4}{3}E_{\mathbf{p}}^2(p_x^2-2p_y^2+p_z^2) \\
\scriptstyle + \frac{\omega^2}{3}(p_x^2+p_y^2+4p_z^2)
\end{array}\cr
\bar{\Delta}^2p_x p_y p_z^2-4E_{\mathbf{p}}^2p_x p_y & \begin{array}{c}
\scriptstyle \bar{\Delta}^2p_y^2 p_z^2+ \frac{4}{3}E_{\mathbf{p}}^2(p_x^2-2p_y^2+p_z^2) \\
\scriptstyle + \frac{\omega^2}{3}(p_x^2+p_y^2+4p_z^2)
\end{array} & 4\bar{\Delta}^2p_y p_z^3-8E_{\mathbf{p}}^2p_y p_z \\
\end{array} \right) \ ,
\eeq
where we now specialize to the ground state given in \Eq{groundState}.

Integration over the spatial momenta eliminates any terms containing odd powers of any of the momenta. The only non-zero contribution is therefore,
\beq
\mathcal{H}[\mathcal{M}^{(1)}] &=& \int \frac{d^3p}{(2\pi)^3}\frac{2\bar{\Delta}\tanh\left(\frac{E_{\mathbf{p}}}{2T}\right)\left[p_y^2 p_z^2\bar{\Delta}^2+ \frac{1}{3}\left(4E_{\mathbf{p}}^2(p_x^2-2p_y^2+p_z^2)+\omega^2(p_x^2+p_y^2+4p_z^2)\right)\right]}{E_{\mathbf{p}}\left(\omega^2+4E_{\mathbf{p}}^2\right)} \mathcal{M}^{(1)}  \propto \mathcal{M}^{(1)} \ .
\eeq
It is straightforward to perform the same operation for the remaining matrices in \Eq{orthBasis}, therefore, these matrices represent the set of normal modes.

\subsection{Angulons}

Because the angulons correspond to rotations about the $x$- and $y$-axes, we may identify them with the generators $J_{1,2} = \mathcal{M}^{(1,2)}$. To show that the angulons remain massless for all temperatures below the critical temperature we need only show that the potential is zero. To do so requires performing the angular integrations, which may only be done analytically before performing the sum over $p_0$. Therefore, we begin with \Eq{S2presum} and set $\omega=0$ and $\delta \Delta =\bar{\Delta} \mathcal{M}^{(1,2)}$ to find the potential. This gives,
\beq
S_2 &=& T\bar{\Delta}^2 \sum_{p_0}\int \frac{d^3p}{(2\pi)^3}  \left[\frac{4(\mathbf{p}\cdot \mathcal{M}^{(1,2)}\cdot\Delta_0\cdot \mathbf{p})^2-2(p_0^2+E_{\mathbf{p}}^2)(\mathbf{p}\cdot\mathcal{M}^{(1,2)}\cdot\mathcal{M}^{(1,2)}\cdot \mathbf{p})}{(p_0^2+E_{\mathbf{p}}^2)^2}+ \frac{\frac{8}{3} (\mathbf{p}\cdot \hat{\Delta}_0\cdot\hat{\Delta}_0\cdot \mathbf{p})}{p_0^2+E_{\mathbf{p}}^2}\right]  \cr
&=& T\bar{\Delta}^2 M k_F^3 \sum_{p_0}\int \frac{d \epsilon dx d\phi}{(2\pi)^3} \left[16(\epsilon^2+p_0^2)(3x^2-1)+4(k_F\bar{\Delta})^2(3x^4+6x^2-1) \right. \cr
&\pm& \left. 12(x^2-1)\left(4\epsilon^2+4p_0^2+(k_F\bar{\Delta})^2(x^2+1)\right)\cos(2\phi)\right]/\left(3\left(4\epsilon^2+4p_0^2+(k_F\bar{\Delta})^2(3x^2+1)\right)^2\right)\cr
&=& \pm T\bar{\Delta}^2 M k_F^3 \sum_{p_0}\int \frac{d \epsilon d\phi}{(2\pi)^3}\frac{16\tan^{-1}\sqrt{\frac{3(k_F \bar{\Delta})^2}{4\epsilon^2+(k_F \bar{\Delta})^2+4p_0^2}}\cos(2\phi)}{3k_F \bar{\Delta}\sqrt{3\left(4\epsilon^2+(k_F \bar{\Delta})^2+4p_0^2\right)}} \ ,  
\eeq
where $x=\cos \theta$, we have used the fact that the integral is dominated by the singularity at $p=k_F$ for small $\bar{\Delta}/v_F$ to approximate $p\approx k_F, \epsilon \approx v_F(p-k_F)$, and the $\pm$ correspond to $\delta \Delta = \mathcal{M}^{(1,2)}$, respectively. 

The remaining integration over the angle $\phi$ is clearly $0$ for both modes, resulting in a flat potential for the angulons. This statement is a consequence of the geometry of the ground state, depending only on the angular integration, and is completely independent of any assumptions about $p_0$. Therefore, the angulons must be massless not only at zero temperature, but for all temperatures below $T_c$. This result is in apparent disagreement with that in ref.~\cite{Leinson:2012pn}. We do not understand the origin of this discrepancy.

\subsection{Massive modes}
To calculate the masses of the remaining modes, we take $\omega \neq 0$ and look for poles of the propagators corresponding to each mode. The inverse propagator for the $i$th mode, corresponding to $\omega_i$, is given by substituting $\delta \Delta = \mathcal{M}^{(i)}$ into \Eq{S2postSum}. 

\begin{figure}
\centering
\includegraphics[width=0.6\linewidth]{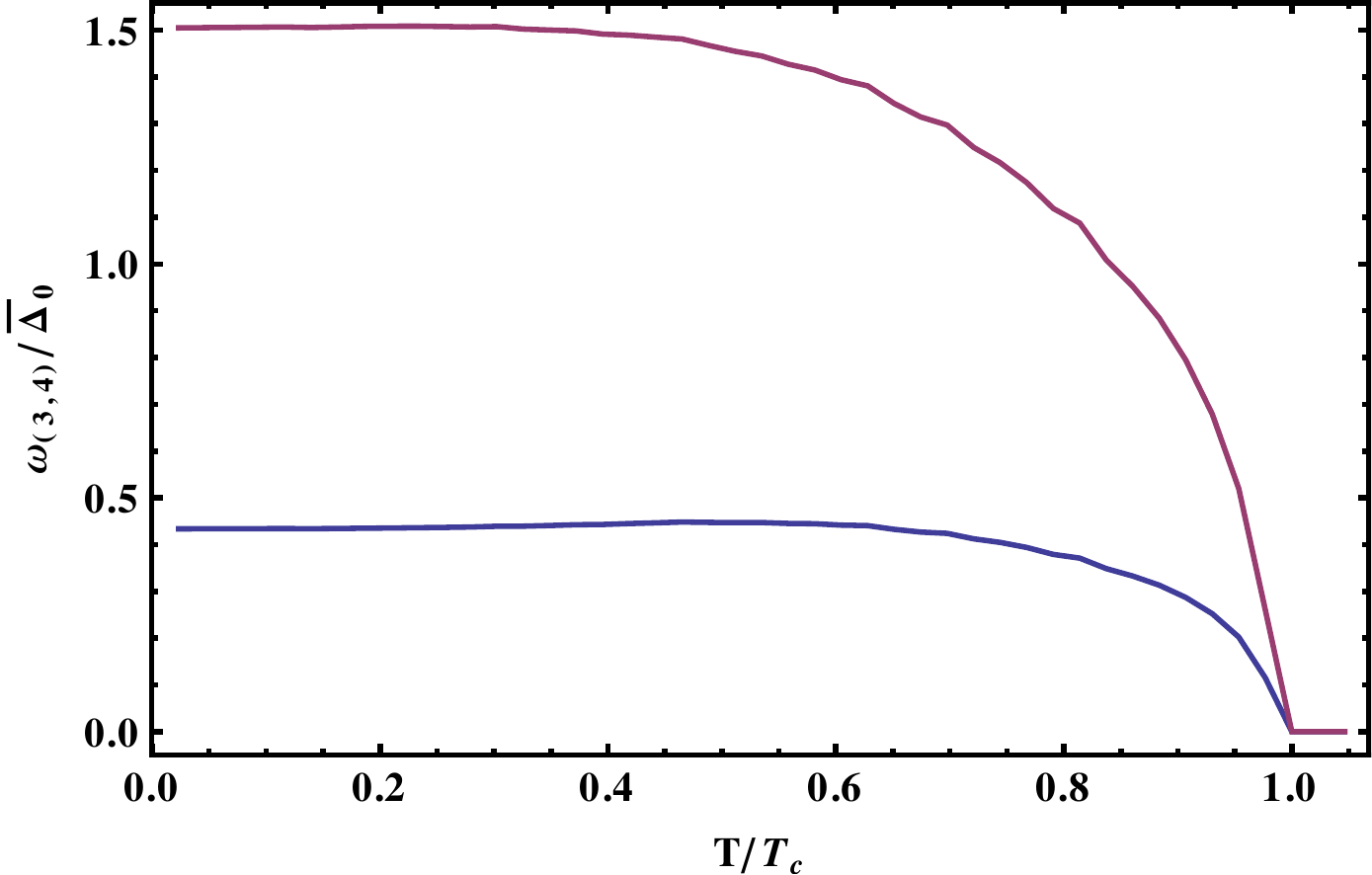}
\caption{\label{fig:masses}Masses of the modes associated with $\mathcal{M}_{(3,4)}$ in units of the magnitude of the gap at zero temperature as a function of the temperature in units of the critical temperature. The two curves represent the two poles found for a given state, and each curve corresponds to two degenerate states.}
\end{figure}

We find that the modes corresponding to $\mathcal{M}^{(3,4)}$ are degenerate. This is due to the remaining symmetry in the $(x,y)$-plane resulting from our choice for the ground state (\Eq{groundState}). The inverse propagator for these modes is given by,
\beq
\Pi_{(3,4)}^{-1}(\omega_{(3,4)}) &=& \bar{\Delta}^2 \frac{M k_F^3}{4\pi^2} \int d\epsilon dx \tanh\left(\frac{E}{2T}\right)\left[ \frac{\left(\frac{(k_F\bar{\Delta})^2 }{2}(1-x^2)^2+\omega_{(3,4)}^2(1-x^2)\right)}{E(4E^2+\omega_{(3,4)}^2)} +  \frac{2(x^2-\frac{1}{3})}{E} \right] \ ,
\eeq
where $E = \sqrt{\epsilon^2 + (k_F \bar{\Delta})^2 (1+3x^2)/4}$. To find the poles we must rotate to Minkowski space $(\omega \to i \omega)$ and numerically integrate. We find two solutions for the equation $\Pi_{(3,4)}^{-1}(i\omega_{(3,4)})  = 0$. Numerical solutions for $\omega_{(3,4)}$ are plotted in \Fig{masses}.

The masses of both modes are below the threshold for pair breakup.  They are also of the order of the gap for low temperatures $T \ll T_c$, therefore, properties of the angulons based on an effective theory valid for energies much smaller than the gap will not be greatly affected by these modes. This is one of the results we initially sought in order to validate the approach of ref. ~\cite{Bedaque:2012bs}. This result seems to be in rough quantitative agreement with ref.~\cite{Leinson:2012pn}.

The inverse propagator corresponding to $\mathcal{M}_5$ is,
\beq
\Pi_{5}^{-1}(\omega_{5}) &=& \bar{\Delta}^2 \frac{M k_F^3}{4\pi^2} \int d\epsilon dx \tanh\left(\frac{E}{2T}\right)\left[ \frac{\left((k_F\bar{\Delta})^2(1+3x^2)+\omega_5^2\right)\left(\frac{1+3x^2}{4}\right)}{E(4E^2+\omega_{5}^2)} \right] \ .
\eeq
We find no solutions to the equation $\Pi_{5}^{-1}(i\omega_{5}) =0$, therefore, there are no modes associated with $\mathcal{M}_5$, again in agreement with ref.~\cite{Leinson:2012pn}.

\begin{figure}[H!]
\centering
\includegraphics[width=0.6\linewidth]{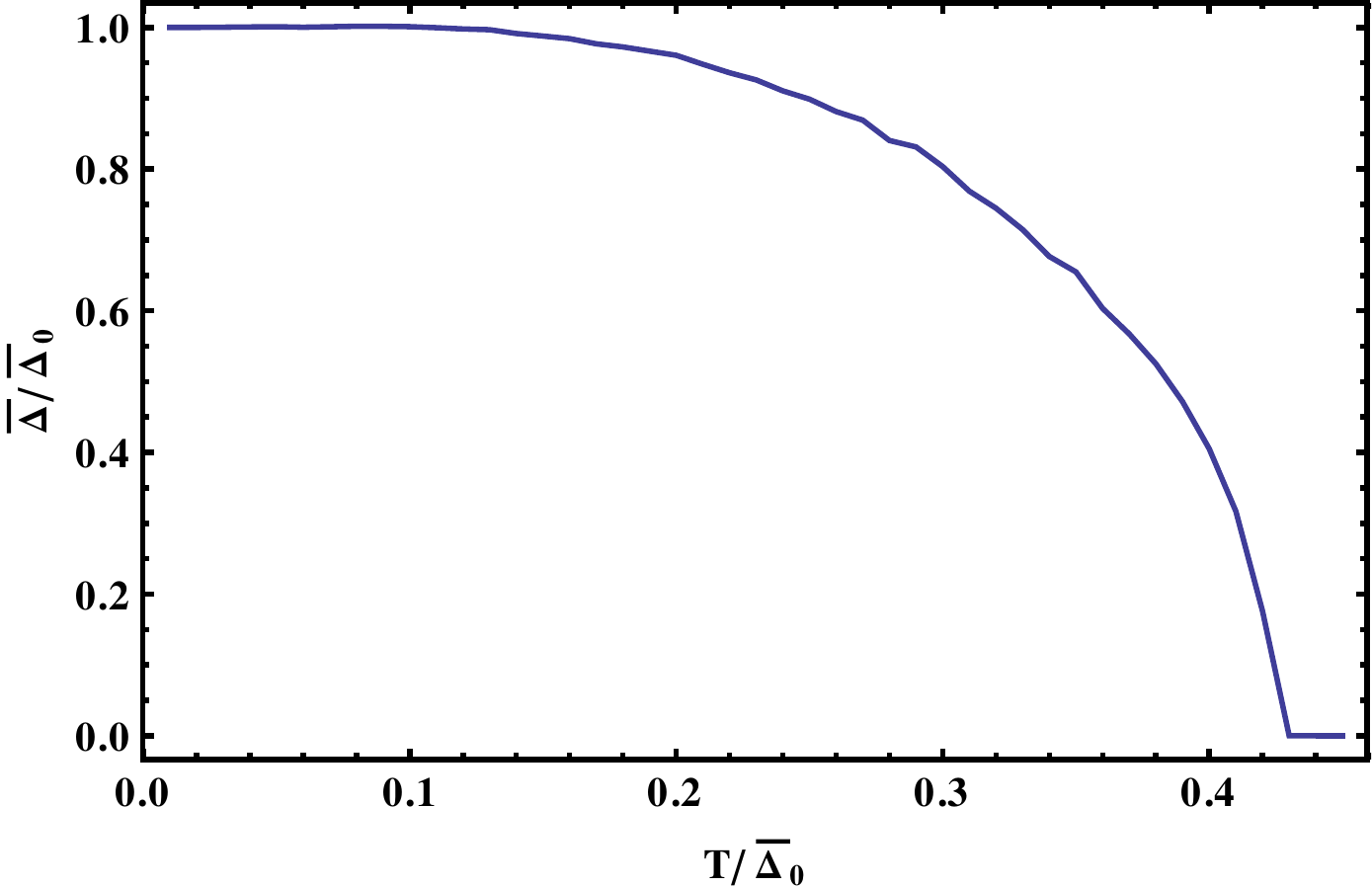}
\caption{\label{fig:gap}Magnitude of the gap as a function of temperature, in units of the magnitude of the gap at zero temperature.}
\end{figure}

\section{Summary}
We have found the spectrum of bosonic modes for \threeptwo condensed neutron matter using a simple model calculation. We find the existence of two massless Goldstone modes associated with spontaneously broken rotational symmetry in two planes for all temperatures below the critical temperature. This is in contrast to the result found in \cite{Leinson:2012pn}, where these modes acquire masses for all nonzero temperatures. In addition, we find two  massive modes whose masses are of the order of the (zero-temperature) gap as long as $T\ll T_c$. The fact that the massive modes have a minimum energy of the order of the zero temperature gap justifies the use of the effective theory developed in ref.~\cite{Bedaque:2012bs}. Due to their large masses, their contributions to processes at temperatures $T\ll T_c$ are exponentially suppressed. For this reason we did not compute their properties at non-zero spatial momentum. The contributions from the massless modes to neutron star physics should be much more relevant and were discussed in ref.~\cite{Bedaque:2013fja}. We have not considered modes corresponding to complex (as was done in \cite{Leinson:2012pn}), non-symmetric, or non-zero trace deformations of the condensate, the latter two corresponding to an admixture of ${}^3P_0$ and ${}^3P_1$ pairing to the ${}^3P_2$ background. There is no reason to expect them to be particularly light, but a calculation of their masses would require the relative strengths of the pairing force in these different channels as an input.

\appendix \section{Gap equation and critical temperature} \label{sec:gap}
The gap equation, $\frac{\delta S}{\delta \Delta}\bigr\rvert_{\Delta=\Delta_0}=0$, gives us,
\beq
\frac{\Delta_{ij}}{2g^2} = 2T \sum_{p_0} \int \frac{d^3p}{(2\pi)^3} \frac{(\mathbf{p} \cdot \Delta)_i \mathbf{p}_j+(\mathbf{p}\cdot \Delta)_j\mathbf{p}_i -\frac{2}{3} \mathbf{p}\cdot \Delta \cdot \mathbf{p} \delta_{ij}}{p_0^2+\epsilon^2+\mathbf{p} \cdot \Delta^2\cdot \mathbf{p}} \ .
\eeq
This leads to the following equation for the magnitude of the gap, using our chosen ground state (\Eq{groundState}),
\beq
\frac{\bar{\Delta}}{4g^2}= \frac{2Tk_F^2M}{3\pi} \sum_{p_0} \int_{-1}^1 dx \frac{(1+3x^2)/4}{\left(\left(p_0/(k_F \bar{\Delta})\right)^2+(1+3x^2)/4\right)^{1/2}} \ ,
\eeq
where $x=\cos \theta$ and we have used the fact that the integral is dominated by the singularity at $p=k_F$ for small $\bar{\Delta}/v_F$ to approximate $p\approx k_F, \epsilon \approx v_F(p-k_F)$. In \Fig{gap} we plot the value of the gap as a function of temperature. We find the critical temperature $T_c \approx 0.43 \bar{\Delta}_0$, where $\bar{\Delta}_0$ is the magnitude of the gap at zero temperature.

\appendix \section{Real time formalism}\label{sec:realTime}

The imaginary time formalism used in this work is only one out of many methods used in studying field theories at finite temperature. It allows for the computation of field correlators using imaginary time and (anti-)periodic boundary conditions in the Euclidean ``time" direction. When information about real time correlators is required an analytic continuation must be made from imaginary and discretized energies $k_0=2\pi i n, n\in \mathbb{Z}$ to real, continuous ones. In general this continuation is not unique, but for two-point functions the correct behavior of the correlator at  asymptotically large $|k_0|$ does specify a unique continuation\cite{baym_mermin}. In practice, this continuation may be difficult to find. In order to verify that we have the correct analytic continuation, we have repeated the calculation described in the main text using the real time formalism (RTF) for finite temperature field theories \cite{feynman1963theory,schwinger,keldysh1965diagram,craig} (for a review see, for example, ref. \cite{Chou19851}). In the RTF the number of fields is doubled and each copy is denoted by an index ``+" or ``-". Propagators acquire a $2\times 2$ matrix structure corresponding to the doubling of the number of fields. The construction of the perturbation series follows the usual diagrammatic rules familiar to the zero temperature case with the addition of vertices involving the ``-" fields (which come with an opposite sign). Fortunately, in our calculation only the ``+" fields appear. The ``++" component of propagator of the fermions is given by
\be
D^{++}(p) = \left[
\frac{1}{p_0^2-E_p^2+i\epsilon} + i 2\pi n(p) \delta(p_0^2-E_p^2)
\right]
\begin{pmatrix}
p_0+\epsilon_p  & (\mathbf{p}. \Delta_0)_j\sigma_i\sigma_2\\
 ( \Delta_0.\mathbf{p})_j\sigma_2\sigma_i\ & p_0-\epsilon_p  
\end{pmatrix},
\ee with
\be
n(p)=\frac{1}{e^{\beta|p_0|}+1}.
\ee The $2\times 2$ structure of the propagator above refers to ``Gorkov space", not the doubling of fields due to the RTF. Repeating the steps in the main text, now in the RTF, we have
\beq
\frac{\delta^2\Tr\log(D^{-1})}{\delta\Delta_{ij}(s)\delta\Delta_{kl}(r)}|_{\Delta=\Delta_0} &=&
\int \frac{d^4p}{(2\pi)^4}\left[D^{++}_{11}(p+k) D^{++}_{22}(k)+D^{++}_{22}(p+k) D^{++}_{11}(k)+D^{++}_{12}(p+k) D^{++}_{21}(k)+D^{++}_{21}(p+k) D^{++}_{12}(k)
\right]\nn\\
&=&
\int \frac{d^4p}{(2\pi)^4} \times \left(8\mathbf{p}_i[\Delta_0\cdot \mathbf{p}]_j\mathbf{p}_k[\Delta_0\cdot \mathbf{p}]_l-4\mathbf{p}_i \mathbf{p}_l\left(p_0(p_0+\omega)+E_{\mathbf{p}}^2\right)
\right]) \delta(s+r)\nn\\
&&\left[\frac{1}{(p_0^2-E_{\mathbf{p}}^2+i\epsilon)} + 2\pi i n(p) \delta(p_0^2-E_{\mathbf{p}}^2)\right]  
\left[ \frac{1}{(p_0+\omega)^2-E_{\mathbf{p}}^2+i\epsilon}+ + 2\pi i n(p+k) \delta((p_0+k_0)^2-E_{\mathbf{p}}^2) \right] \nn\\.
\eeq Separating the real from the imaginary part using
\be
\frac{1}{x+i\epsilon} = \frac{x}{x^2+\epsilon^2}-i\frac{\epsilon}{x^2+\epsilon^2} = \mathcal{P}\left( \frac{1}{x}\right) -  i\pi \delta(x),
\ee performing the $p_0$ integral and adding the contribution from the $\Delta_{ij}^\dagger\Delta_{ji}/(4 g^2)$ term we find for the real part of the action (for $\omega\neq 0$)

\beq
S^R_2 =\int \frac{d^3p}{(2\pi)^3}\tanh\left(\frac{E_{\mathbf{p}}}{2T}\right)\left[\frac{4(\mathbf{p}\cdot \delta\Delta\cdot\Delta_0\cdot \mathbf{p})^2-4E_{\mathbf{p}}^2(\mathbf{p}\cdot \delta\Delta\cdot\delta\Delta\cdot \mathbf{p})}{E_{\mathbf{p}}\left(\omega^2-4E_{\mathbf{p}}^2\right)} + \frac{\frac{2}{3}\Tr [\delta \Delta \cdot \delta \Delta]  (\mathbf{p}\cdot \hat{\Delta}_0\cdot\hat{\Delta}_0\cdot \mathbf{p})}{E_{\mathbf{p}}}\right]\ .
\eeq In agreement with eq.~\ref{eq:S2postSum}. The imaginary part, which we do not compute here, describes the thermal width of the quasi-particles. 
%

\begin{acknowledgments}
This work was supported in part by U.S.\ DOE grant No.\ DE-FG02-93ER-40762.
\end{acknowledgments}
\bibliography{3P2ref}

\begin{thebibliography}{13}
\expandafter\ifx\csname natexlab\endcsname\relax\def\natexlab#1{#1}\fi
\expandafter\ifx\csname bibnamefont\endcsname\relax
  \def\bibnamefont#1{#1}\fi
\expandafter\ifx\csname bibfnamefont\endcsname\relax
  \def\bibfnamefont#1{#1}\fi
\expandafter\ifx\csname citenamefont\endcsname\relax
  \def\citenamefont#1{#1}\fi
\expandafter\ifx\csname url\endcsname\relax
  \def\url#1{\texttt{#1}}\fi
\expandafter\ifx\csname urlprefix\endcsname\relax\def\urlprefix{URL }\fi
\providecommand{\bibinfo}[2]{#2}
\providecommand{\eprint}[2][]{\url{#2}}

\bibitem[{\citenamefont{Bedaque et~al.}(2003)\citenamefont{Bedaque, Rupak, and
  Savage}}]{Bedaque:2003wj}
\bibinfo{author}{\bibfnamefont{P.~F.} \bibnamefont{Bedaque}},
  \bibinfo{author}{\bibfnamefont{G.}~\bibnamefont{Rupak}}, \bibnamefont{and}
  \bibinfo{author}{\bibfnamefont{M.~J.} \bibnamefont{Savage}},
  \bibinfo{journal}{Phys.Rev.} \textbf{\bibinfo{volume}{C68}},
  \bibinfo{pages}{065802} (\bibinfo{year}{2003}), \eprint{nucl-th/0305032}.

\bibitem[{\citenamefont{Bedaque and Nicholson}(2013)}]{Bedaque:2012bs}
\bibinfo{author}{\bibfnamefont{P.~F.} \bibnamefont{Bedaque}} \bibnamefont{and}
  \bibinfo{author}{\bibfnamefont{A.~N.} \bibnamefont{Nicholson}},
  \bibinfo{journal}{Phys.Rev.} \textbf{\bibinfo{volume}{C87}},
  \bibinfo{pages}{055807} (\bibinfo{year}{2013}), \eprint{1212.1122}.

\bibitem[{\citenamefont{Bedaque and Reddy}(2013)}]{Bedaque:2013fja}
\bibinfo{author}{\bibfnamefont{P.~F.} \bibnamefont{Bedaque}} \bibnamefont{and}
  \bibinfo{author}{\bibfnamefont{S.}~\bibnamefont{Reddy}}
  (\bibinfo{year}{2013}), \eprint{1307.8183}.

\bibitem[{\citenamefont{Bedaque and Sen}(2014)}]{Bedaque:2013rya}
\bibinfo{author}{\bibfnamefont{P.}~\bibnamefont{Bedaque}} \bibnamefont{and}
  \bibinfo{author}{\bibfnamefont{S.}~\bibnamefont{Sen}},
  \bibinfo{journal}{Phys.Rev.} \textbf{\bibinfo{volume}{C89}},
  \bibinfo{pages}{035808} (\bibinfo{year}{2014}), \eprint{1312.6632}.

\bibitem[{\citenamefont{Leinson}(2012)}]{Leinson:2012pn}
\bibinfo{author}{\bibfnamefont{L.}~\bibnamefont{Leinson}},
  \bibinfo{journal}{Phys.Rev.} \textbf{\bibinfo{volume}{C85}},
  \bibinfo{pages}{065502} (\bibinfo{year}{2012}), \eprint{1206.3648}.

\bibitem[{\citenamefont{Richardson}(1972)}]{Richardson:1972xn}
\bibinfo{author}{\bibfnamefont{R.}~\bibnamefont{Richardson}},
  \bibinfo{journal}{Phys.Rev.} \textbf{\bibinfo{volume}{D5}},
  \bibinfo{pages}{1883} (\bibinfo{year}{1972}).

\bibitem[{\citenamefont{Vulovic and Sauls}(1984)}]{Vulovic:1984kc}
\bibinfo{author}{\bibfnamefont{V.}~\bibnamefont{Vulovic}} \bibnamefont{and}
  \bibinfo{author}{\bibfnamefont{J.}~\bibnamefont{Sauls}},
  \bibinfo{journal}{Phys.Rev.} \textbf{\bibinfo{volume}{D29}},
  \bibinfo{pages}{2705} (\bibinfo{year}{1984}).

\bibitem[{\citenamefont{Gordon~Baym}(19612)}]{baym_mermin}
\bibinfo{author}{\bibfnamefont{D.~N.~M.} \bibnamefont{Gordon~Baym}},
  \bibinfo{journal}{J. Math. Phys.} p. \bibinfo{pages}{232}
  (\bibinfo{year}{19612}).

\bibitem[{\citenamefont{Feynman and Vernon~Jr}(1963)}]{feynman1963theory}
\bibinfo{author}{\bibfnamefont{R.~P.} \bibnamefont{Feynman}} \bibnamefont{and}
  \bibinfo{author}{\bibfnamefont{F.}~\bibnamefont{Vernon~Jr}},
  \bibinfo{journal}{Annals of physics} \textbf{\bibinfo{volume}{24}},
  \bibinfo{pages}{118} (\bibinfo{year}{1963}).

\bibitem[{\citenamefont{Schwinger}(1961)}]{schwinger}
\bibinfo{author}{\bibfnamefont{J.}~\bibnamefont{Schwinger}},
  \bibinfo{journal}{J. Math. Phys.} \textbf{\bibinfo{volume}{2}},
  \bibinfo{pages}{407} (\bibinfo{year}{1961}).

\bibitem[{\citenamefont{Keldysh}(1965)}]{keldysh1965diagram}
\bibinfo{author}{\bibfnamefont{L.}~\bibnamefont{Keldysh}},
  \bibinfo{journal}{Sov. Phys. JETP} \textbf{\bibinfo{volume}{20}},
  \bibinfo{pages}{1018} (\bibinfo{year}{1965}).

\bibitem[{\citenamefont{Craig}(1968)}]{craig}
\bibinfo{author}{\bibfnamefont{R.~A.} \bibnamefont{Craig}},
  \bibinfo{journal}{J. Math. Phys.} \textbf{\bibinfo{volume}{9}},
  \bibinfo{pages}{605} (\bibinfo{year}{1968}).

\bibitem[{\citenamefont{chao Chou et~al.}(1985)\citenamefont{chao Chou, bin Su,
  lin Hao, and Yu}}]{Chou19851}
\bibinfo{author}{\bibfnamefont{K.}~\bibnamefont{chao Chou}},
  \bibinfo{author}{\bibfnamefont{Z.}~\bibnamefont{bin Su}},
  \bibinfo{author}{\bibfnamefont{B.}~\bibnamefont{lin Hao}}, \bibnamefont{and}
  \bibinfo{author}{\bibfnamefont{L.}~\bibnamefont{Yu}},
  \bibinfo{journal}{Physics Reports} \textbf{\bibinfo{volume}{118}},
  \bibinfo{pages}{1 } (\bibinfo{year}{1985}), ISSN \bibinfo{issn}{0370-1573},
  \urlprefix\url{http://www.sciencedirect.com/science/article/pii/037015738590%
136X}.

\end{thebibliography}
\end{document}